\documentclass[12pt]{article}
\usepackage{amssymb,amsmath,bm}
\usepackage{epsf}
\usepackage{epsfig}
\usepackage{color}
\usepackage{longtable}

\def\simge{\mathrel{\rlap{\raise 0.511ex \hbox{$>$}}{\lower 0.511ex \hbox{$\sim$}}}}
\def\simle{\mathrel{\rlap{\raise 0.511ex \hbox{$<$}}{\lower 0.511ex \hbox{$\sim$}}}}
\def\slash#1{\setbox0=\hbox{$#1$}\dimen0=\wd0
      \setbox1=\hbox{/} \dimen1=\wd1 \ifdim\dimen0>\dimen1
      \rlap{\hbox to \dimen0{\hfil/\hfil}} #1                        \else
      \rlap{\hbox to \dimen1{\hfil$#1$\hfil}}
      /   \fi}

\newcommand{\lsim}{
\mathrel{\hbox{\rlap{\hbox{\lower4pt\hbox{$\sim$}}}\hbox{$<$}}}}

\newcommand{\gsim}{
\mathrel{\hbox{\rlap{\hbox{\lower4pt\hbox{$\sim$}}}\hbox{$>$}}}}

\newcommand{\tev}{\, {\rm TeV}}
\newcommand{\gev}{\, {\rm GeV}}

\newcommand{\Heff}{{\cal H}_{\rm eff}}

\allowdisplaybreaks[2]

\newcommand{\be}{\begin{equation}}
\newcommand{\ee}{\end{equation}}
\newcommand{\bea}{\begin{eqnarray}}
\newcommand{\eea}{\end{eqnarray}}
\newcommand{\nn}{\nonumber}
\newcommand{\bi}{\begin{itemize}}
\newcommand{\ei}{\end{itemize}}
\newcommand{\ord}{{\cal O}}

\newcommand{\newsection}[1]{\section{#1}\setcounter{equation}{0}}

\begin{document}
\begin{titlepage}
\vspace*{-0.5truecm}

%{\bf\large\today}

\begin{flushright}
TUM-HEP-648/06\\
MPP-2006-124
\end{flushright}

\vfill

\begin{center}
\boldmath

{\Large\textbf{Lower Bounds on $\Delta M_{s,d}$ from \vspace{0.2cm}\\Constrained
    Minimal Flavour Violation}}

\unboldmath
\end{center}

\vspace{0.4truecm}

\begin{center}
{\large\bf Monika Blanke$^{a,b}$ and Andrzej J.~Buras$^a$}
\vspace{0.4truecm}

 {\sl $^a$Physik Department, Technische Universit\"at M\"unchen,
D-85748 Garching, Germany}
\vspace{0.2cm}

 {\sl $^b$Max-Planck-Institut f{\"u}r Physik (Werner-Heisenberg-Institut), \\
D-80805 M{\"u}nchen, Germany}

\end{center}

\vspace{1cm}
\begin{abstract}
\vspace{0.2cm}
\noindent 
We demonstrate that in the class of models with Constrained Minimal
Flavour Violation (CMFV), in which the CKM matrix is the only source
of CP violation and $B^0_q-\bar B^0_q$ mixings $(q=d,s)$ are described
by a single operator $(\bar b q)_{V-A} (\bar b q)_{V-A}$, the lower
bounds on the mass differences $\Delta M_{s,d}$ are simply given by
their Standard Model values $(\Delta M_{s,d})_\text{SM}$ with two possible exceptions that we identify. Our proof involves all possible charged gauge boson, Goldstone boson and physical charged and neutral scalar exchanges and
assumes that the masses of new charge $+2/3$ singlet heavy fermions
$T_i$ that mix with the top quark have all to be larger than $m_t$,
which is chosen by nature. Similarly, the additional charged gauge bosons have to satisfy $M_i>M_W$, while no such bound needs to be set on the charged and neutral scalar masses. The two possible exceptions arise in the presence of $U(1)$ neutral gauge bosons and Majorana fermions in box diagrams with flavour violating couplings. {However in the case of the MSSM {with CMFV and} low $\tan\beta$ our bound is satisfied in spite of gluino and neutralino contributions.}
\end{abstract}
\vfill\vfill\vfill
\end{titlepage}

\setcounter{page}{1}
\pagenumbering{arabic}

\newsection{Introduction}
\label{sec:intro}

The $B^0_{d,s} - \bar B^0_{d,s}$ mixing mass differences $\Delta
M_{d,s}$ played, in the last 20 years, a prominent role in the
determination of the CKM matrix \cite{CKM}. Simultaneously, being
generated first at the one-loop level in weak interactions, $\Delta
M_{d,s}$ played an important role in the tests of the short distance
structure of the Standard Model (SM) and of its extensions, like the
MSSM, models with Extra Dimensions, multi-Higgs models and Little
Higgs models with and without T-parity.

With the advent of tree level methods for the determination of the CKM
matrix \cite{CPreview}, that, in contrast to $\Delta M_{d,s}$, are to a
very good approximation independent of new physics contributions, the
role of $\Delta M_{d,s}$ will be shifted primarily to the tests of the
SM and its extensions.

One of the simplest classes {of extensions of the SM} are models
with Constrained Minimal Flavour Violation (CMFV) \cite{CMFV} in which\footnote{A more general formulation of Minimal Flavour Violation can be found in \cite{mfv2}. Our present discussion applies only to CMFV as formulated in \cite{CMFV} and here.}
{
\begin{enumerate}
\item
the SM Yukawa couplings and the resulting CKM matrix are the only source of flavour {and CP} violation,
\item only operators are relevant that are relevant already in the SM.
\end{enumerate}
Consequently,}
$\Delta M_{d,s}$ are governed by the CKM factors
\be
\lambda_t^{(d)}=V_{tb}^* V_{td}^{} \,,\qquad
\lambda_t^{(s)}=V_{tb}^* V_{ts}^{}\,,
\ee
and the single local operator in the effective theory is $(q=d,s)$
\be\label{eq:Qq}
\mathcal{Q}_q=(\bar b \gamma_\mu(1-\gamma_5)q)\otimes(\bar b
\gamma^\mu(1-\gamma_5)q)\,.
\ee
Therefore, in this class of models one simply has
\cite{LesHouches}
\be\label{eq1.3}
\Delta M_q = \frac{G_F^2}{6\pi^2}\eta_B m_{B_q}\hat B_{B_q} F_{B_q}^2
M_W^2 |S(y)| |\lambda_t^{(q)}|^2\,,
\ee
where $\hat B_{B_q} F_{B_q}^2$ result from the non-perturbative
evaluation of the relevant hadronic matrix elements of $\mathcal{Q}_q$, and
$\eta_B$ is the QCD correction that in the SM is found at the NLO
level to be $\eta_B=0.55$ \cite{etaB}.

The most important quantity in the present paper is the short distance
function 
\be
S(y)=S_0(x_t)+\Delta S(y)\,,
\ee
resulting from box diagrams. Here $S_0(x_t)$ with $x_t=m_t^2/M_W^2$, introduced in \cite{S0}, is
one of the so-called Inami-Lim functions \cite{InamiLim} in the SM, and
$\Delta S(y)$ summarizes all new physics contributions in a given CMFV
model with $y$ denoting collectively the parameters of this model. It
should be emphasized that all CP-violating phases are collected in
this class of models in $\lambda_t^{(q)}$ so that $S(y)$ is a real
function. The explicit expression for $S_0(x_t)$ can be found in
\cite{LesHouches}. For the $\overline{\text{MS}}$ top quark mass
$\overline{m}_t=(161.7\pm2.0)\gev$ one has $S_0(x_t)=2.28\pm0.04$.

As already emphasized in \cite{BBGT}, in all known CMFV models,
$\Delta S(y)$ turns out to be positive so that $\Delta M_q > (\Delta
M_q)_\text{SM}$ appears to be a property of this class of models. This
is the case for Two-Higgs-Doublet Models type II \cite {2HDM},
the MSSM with {CMFV} and low $\tan\beta$ \cite{MSSMlow,MSSMlow2}, the Littlest Higgs
model (LH) without T-parity \cite{BPU,Inder}, the CMFV limit of the LH
model with T-parity \cite{BBPTUW} and Universal Extra Dimensions \cite{UED}.  {Concerning the MSSM with {CMFV} and low $\tan\beta$ analyzed in \cite{MSSMlow,MSSMlow2}, we would like to warn the reader{, following \cite{ABG},} that these papers did not include gluino {and neutralino} contributions, and consequently the issue of negative contributions due to these particles being exchanged has not been addressed there. We will return to this issue in Section \ref{sec:neutral}.}

On the other hand, in the MSSM with MFV and large $\tan\beta$, where new operators contribute to $\Delta M_{s,d}$, $\Delta M_s < (\Delta M_s)_\text{SM}$ in the full space of the parameters of the model \cite{BCRS}. In more complicated models, like the MSSM with new flavour violating interactions \cite{ref14}, or the LH model with T-parity \cite{BBPTUW}, $\Delta M_s$ can be smaller or larger than $(\Delta M_s)_\text{SM}$.

On the experimental side both mass differences are known with high
precision \cite{hfag,CDFD0} 
\be
\Delta M_d = (0.508\pm 0.004)/\text{ps}\,,\qquad
\Delta M_s = (17.77\pm0.10\pm 0.07)/{\rm ps}
\ee
with the latter measurement being one of the highlights of 2006. The SM is compatible with both values, even if the predictions for $(\Delta M_s)_\text{SM}$ prior to the measurements displayed the central values in the ballpark of $21/$ps \cite{CKMfitUTfit}. Unfortunately, due to sizable non-perturbative uncertainties in $\hat B_{B_q} F_{B_q}^2$ \cite{BF06}, it will still take some time before we will know through the direct calculation of \eqref{eq1.3} whether $\Delta M_s > (\Delta M_s)_\text{SM}$ or $\Delta M_s < (\Delta M_s)_\text{SM}$ or even $\Delta M_s \simeq (\Delta M_s)_\text{SM}$ with high precision. Some proposals for clarifying this issue with the help of the CP-violating asymmetry $S_{\psi\phi}$ and the semi-leptonic CP asymmetry $A^{s}_\text{SL}$ have been put forward in \cite{BBGT}, but they are very challenging as well.

Now, as stated above, in all CMFV models studied in the literature, $\Delta M_{d,s} > (\Delta M_{d,s})_\text{SM}$ provided the CKM parameters have been determined by means of tree level decays, i.\,e. independently of new physics. The question then arises whether one could prove this to be a general property of the CMFV class by assuming
\begin{enumerate}
\item
three generations of the standard quarks (intrinsic property of CMFV anyway),
\item
GIM mechanism \cite{GIM}, even in the presence of new heavy singlet
$T_i$ quarks with electric charge $+2/3$ that mix only with the standard top quark.
\end{enumerate}

The main goal of the present paper is to demonstrate diagrammatically that indeed $(\Delta M_{d,s})_\text{SM}$ is the lower bound for $\Delta M_{d,s}$ in the class of models with CMFV as defined in \cite{CMFV} that satisfy the two properties listed above, provided only charged gauge bosons, charged and neutral scalars and charged fermions are present in the box diagrams, and mild conditions on the spectrum of the gauge bosons and fermions are satisfied. The bound is also valid for tree-level neutral gauge boson exchanges, while tree-level scalar exchanges give different operators and are outside the CMFV framework. On the other hand we find that the presence of $U(1)$ neutral gauge bosons and Majorana fermions\footnote{We thank Diego Guadagnoli and Gino Isidori for bringing this case to our attention.} 
in the box diagrams with flavour violating couplings could in principle violate our bound. The neutral gauge boson contributions are strictly negative, but we are not aware of any model with neutral gauge boson contributions in box diagrams that could be put into the CMFV class. {On the other hand, in the MSSM at low $\tan\beta$ with universal soft breaking terms introduced at a scale much higher than $M_W$, the presence of gluinos in box diagrams could in principle violate our bound.} {However a recent detailed analysis of the MFV limit of the MSSM at low $\tan\beta$ shows that this does not happen \cite{ABG}. {Moreover, as pointed out in that paper, the MSSM with MFV and low $\tan\beta$ does  not belong to the CMFV class, if the higgsino coupling is large compared to the other SUSY scales.}    We will elaborate on this in Section \ref{sec:neutral}.}

Our paper is organized as follows. In Section \ref{sec:bosons} we will consider contributions from charged gauge bosons, Goldstone bosons and physical charged scalars. { In Section \ref{sec:GF} we discuss how the measured Fermi constant can be modified by new physics contributions and study the implications on the predicted values for $\Delta M_{d,s}$.} In Section \ref{sec:Ti} we study the contributions of an arbitrary number of heavy singlet $T_i$ quarks that mix with the top quark. {In Section \ref{sec:general} we show that the above cases cover indeed all charged particle contributions possible within the CMFV framework.} In Section \ref{sec:neutral} we address the issue of neutral particle exchanges (gauge bosons, fermions and scalars) advertised above. 
We briefly summarize our work in Section \ref{sec:summary}.

\newsection{Contributions from charged Gauge Bosons and Scalars}
\label{sec:bosons}

For our purposes it will be sufficient to consider the three box diagrams shown in Fig.~1 and calculate them in the 't Hooft-Feynman gauge.
The first diagram, $D_1$, describes the contribution of two gauge bosons with masses $M_1$ and $M_2$ and ordinary up-type quarks. The second diagram, $D_2$, corresponds to two scalar exchanges that can be physical scalars and Goldstone bosons. Finally, the third diagram, $D_3$, gives mixed gauge and scalar contributions.  

\begin{figure}
\center{\epsfig{file=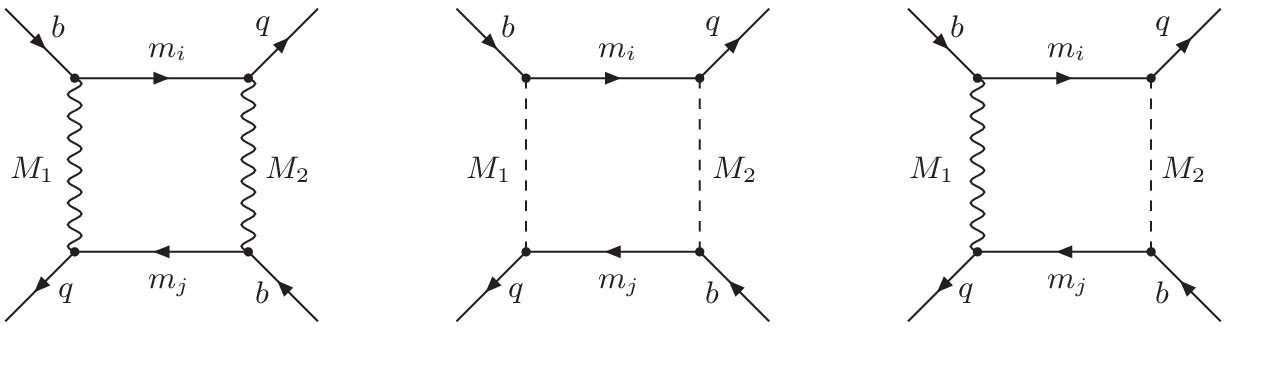}}
\caption{\it Classes of diagrams contributing to $\Delta B=2$ processes in
  CMFV models.}
\end{figure}

In giving the formulae below we will simply multiply the result of
each diagram by a factor of $i$ to obtain its contribution to the
effective Hamiltonian for $\Delta B=2$ transitions, which takes the general form
\be\label{eq:Heff}
\Heff(\Delta B=2)=c \left( \lambda_t^{(q)}\right)^2 S(y) \mathcal{Q}_q\,,
\ee
with $c$ being a positive constant that includes for instance $G_F$ and QCD corrections. We recall that 
\be
\Delta M_q = \frac{\left|\langle\bar B^0_q |\Heff(\Delta B=2) | B^0_q\rangle\right|}{m_{B_q}}\,.
\ee

In what follows we will not
include combinatorial factors like a factor of 2 for two different
bosons being exchanged, nor QCD corrections, as they do not have any impact on our basic result. 
Indeed, as we will demonstrate below, each of the diagrams in Fig.~1 gives a positive contribution to $\Heff$ in \eqref{eq:Heff} times the relevant CKM factor. Taking next into account the contributions from ordinary $u,c,t$ quarks, with $m_u,m_c \ll m_t$, relevant for the $\lambda_t^{(q)}$ term in \eqref{eq1.3}, and using the unitarity of the CKM matrix, we show that the known GIM structure in the case of $\Delta B=2$ processes
\be\label{eq2.1}
S(y) \propto F(t,t) + F(u,u) - 2 F(u,t)\,,
\ee
resulting from each diagram in Fig.~1 independently, remains positive,
which completes the proof. Note that by extending the theory with
additional gauge bosons and scalars, the SM part remains unchanged and
the new physics leads simply to additional positive contributions.  Here $F(i,j)$ denotes the short distance
function up to a common overall factor that corresponds to a box
diagram with $i$ and $j$ quark exchanges.

Calculating then the diagrams $D_1$, $D_2$ and $D_3$ of Fig.~1 in the
't Hooft-Feynman gauge, and multiplying them by $i$, we find $(i,j=u,c,t)$
{\bea
i D_1 &=& i \frac{g_1^2
  g_2^2}{16}\lambda_i^{(q)}\lambda_j^{(q)}I_1(m_i,m_j;M_1,M_2)
\mathcal{Q}_q\,,\label{eq2.2}\\
i D_2 &=& i \frac{k_{i1}k_{i2}k_{j1}k_{j2}}{4}\frac{g_1^2
  g_2^2}{16}\lambda_i^{(q)}\lambda_j^{(q)}I_1(m_i,m_j;M_1,M_2)
\mathcal{Q}_q\,,\\
i D_3 &=& -i k_{i2}k_{j2}m_i m_j \frac{g_1^2
  g_2^2}{16}\lambda_i^{(q)}\lambda_j^{(q)}I_2(m_i,m_j;M_1,M_2)
\mathcal{Q}_q\,.\label{eq2.4}
\eea
Here, $g_i$ are the gauge couplings, with $g_1=g_2=g$ in the SM, that in some extensions of the SM
should be replaced in $D_2$ and $D_3$ by other model dependent
couplings, {which however, due to the CMFV hypothesis, always have to be
flavour universal}. The factors $k_{ia}$ $(i=u,c,t;\;a=1,2)$ denote the scalar-fermion couplings. In the SM, they are simply given by  $k_{ia}\propto m_i$. In a general CMFV model they can be written as
\be\label{eq:kia}
k_{ia} = \alpha_a + \beta_a m_i + {\ord(m_i^2)}\,,
\ee
where $\alpha_a,\beta_a\in\mathbb{R}$ are flavour universal constants that may depend on the scalar $a$.
}

Finally, the two integrals $I_1$ and $I_2$ are given as follows
\bea
I_1(m_i,m_j;M_1,M_2)&=&\int\frac{d^4k}{(2\pi)^4}\,\frac{k^2}{(k^2-m_i^2)(k^2-m_j^2)(k^2-M_1^2)(k^2-M_2^2)}\,,\\
I_2(m_i,m_j;M_1,M_2)&=&\int\frac{d^4k}{(2\pi)^4}\,\frac{1}{(k^2-m_i^2)(k^2-m_j^2)(k^2-M_1^2)(k^2-M_2^2)}\,.
\eea

Using the standard techniques, i.\,e. Wick rotation and integration
over the angular directions, they can be
rewritten as follows
\bea\label{eq2.8}
i I_1(m_i,m_j;M_1,M_2)&=&\frac{1}{16\pi^2}\int_0^\infty
dr\,\frac{r^2}{(r+m_i^2)(r+m_j^2)(r+M_1^2)(r+M_2^2)}\,,\\
-i  I_2(m_i,m_j;M_1,M_2)&=&\frac{1}{16\pi^2}\int_0^\infty
dr\,\frac{r}{(r+m_i^2)(r+m_j^2)(r+M_1^2)(r+M_2^2)}\,,\label{eq2.9}
\eea
implying that all three expressions in \eqref{eq2.2}-\eqref{eq2.4}
are, up to CKM factors $\lambda_{i,j}^{(q)}$, real and positive
quantities.

This result remains true for the linear combination in \eqref{eq2.1}
that enters the function $S(y)$ after the CKM unitarity relation
\be\label{eq2.10}
\lambda_u^{(q)}+\lambda_c^{(q)}+\lambda_t^{(q)}=0
\ee
has been used.

In order to see this in the case of pure gauge boson contributions, it
is sufficient to look at the two quark propagators in
\eqref{eq2.8}. Then the linear combination in \eqref{eq2.1} reads
\be\label{eq2.11}
\frac{1}{(r+m_t^2)^2}+\frac{1}{r^2}-\frac{2}{r(r+m_t^2)}=\left(\frac{1}{r}-\frac{1}{r+m_t^2}\right)^2
>0\,.
\ee
{
In the case of $D_2$, we find
\be\label{GIM2}
\frac{(k_{t1}k_{t2})^2}{(r+m_t^2)^2}+\frac{(k_{u1}k_{u2})^2}{r^2}-\frac{2k_{u1}k_{u2}k_{t1}k_{t2}}{r(r+m_t^2)}=\left(\frac{k_{u1}k_{u2}}{r}-\frac{k_{t1}k_{t2}}{r+m_t^2}\right)^2>0\,.
\ee
In the case of $D_3$ the functions $F(u,u)$ and $F(u,t)$
vanish due to $m_u=0$, and as
$F(t,t)>0$, we again obtain positive contributions.}

We have thus shown that, after the GIM mechanism has been used, all three
classes of contributions, represented by the three diagrams in Fig.~1,
give separately positive contributions to the function $S(y)$. This
means that all new physics contributions of this class have the same
sign as the SM contribution, which implies $\Delta S(y)>0$. We have verified that the same result follows in the unitary gauge, where only physical scalars enter the diagrams in Fig.~1. 

{ The above calculation applies not only to the three ordinary quark generations, but also to their heavy partners, that do not mix with the standard ones. This happens e.\,g. in models with Universal Extra Dimensions \cite{UED}, where the heavy partners of the SM particles are the Kaluza-Klein modes of the latter. In this context it is important to note that the Yukawa couplings of the heavy partner fields have to be proportional to the masses of the corresponding SM quarks, in order not to violate CMFV. {Note that the masses $m_im_j$ multiplying $D_3$ now have to be replaced} by the heavy partner masses. A straightforward calculation following the lines of \eqref{eq2.11} and \eqref{GIM2} shows that also in this case the contribution from $D_3$ is positive. 

We would like to remark that this proof also applies to {chargino contributions in} supersymmetric models with {CMFV} and low $\tan\beta$, in which the supersymmetric partners of the SM fields contribute only through diagrams of the type $D_2$. {Note that in this case the GIM mechanism runs over the squarks, i.\,e. the scalars present in the box diagram $D_2$.} {Moreover, the index $a$ in $k_{ia}$ of \eqref{eq:kia} now corresponds to the charginos.} The property of $\Delta S(y) >0$ in the MSSM with {CMFV and low $\tan\beta$ has been pointed out in \cite{MSSMlow} and confirmed in \cite{MSSMlow2}, {but as already stated above these papers considered only Dirac fermion contributions, as we did in this section. The case of Majorana fermions is discussed in Section \ref{sec:neutral}.}

\newsection{Modification of $\bm{G_F}$}\label{sec:GF}

Until now we have assumed that  the SM gauge coupling $g$ is known from experiment. This is however not exactly true. What is measured is the $W^\pm$ boson mass $M_W$ and the Fermi constant $G_F$, obtained from the decay $\mu^-\to e^-\nu_\mu\bar\nu_e$ and given in the SM by 
\be
G_F=\frac{g^2}{4\sqrt{2}M_W^2}\,.
\ee
However, new charged gauge bosons and charged scalars also contribute to this decay, so that the expression for the measured $G_F^\text{eff}$ in terms of $g^2$ gets modified as follows (see \cite{BPU} for a discussion of this issue in the LH model)
\be\label{eqGFeff}
G_F^\text{eff}=\frac{g^2}{4\sqrt{2}M_W^2}+\sum_i\frac{g_i^2}{4\sqrt{2}M_i^2}-\sum_j\frac{{g^\Phi_j}^2}{4\sqrt{2}{M^\Phi_j}^2}\,,
\ee
{implying a modified value of the determined $g^2$.}

Here we have assumed the presence of additional gauge bosons $W^\pm_i$ with masses $M_i$ and gauge couplings $g_i$ and additional physical scalars $\Phi^\pm_j$ with couplings $g^\Phi_j$ and masses $M^\Phi_j$.
We find that additional gauge bosons lead to positive contributions to $G_F^\text{eff}$, while physical scalars tend to lower its value. This is due to the relative minus sign of the scalar propagator with respect to the gauge boson propagator.
{Consequently the presence of scalar contributions to $G_F^\text{eff}$ always implies an increased value of $g^2$ and thus a further enhancement of $\Delta M_{d,s}$.}

So, in the following we will concentrate only on additional gauge bosons present in the theory. In this case, solving \eqref{eqGFeff} for $g^2$, we find
\be
g^2=4\sqrt{2}M_W^2 G_F^\text{eff}-\sum_i\frac{M_W^2}{M_i^2}g_i^2\,.
\ee

Inserting this into the contributions of $W^\pm$ and $W^\pm_i$ to the integrand in \eqref{eq2.8}, \eqref{eq2.9} leads to
\bea
\left(\frac{g^2}{r+M_W^2}+\sum_i \frac{g_i^2}{r+M_i^2}\right)^2
&=&\left(\frac{4\sqrt{2}G_F^\text{eff}M_W^2}{r+M_W^2}-\sum_i\frac{M_W^2}{M_i^2}\frac{g_i^2}{r+M_W^2}+\sum_i\frac{g_i^2}{r+M_i^2}\right)^2\nn\\
&=&\left(\frac{4\sqrt{2}G_F^\text{eff}M_W^2}{r+M_W^2}+\sum_i\frac{g_i^2}{M_i^2}\frac{r(M_i^2-M_W^2)}{(r+M_W^2)(r+M_i^2)}\right)^2.
\eea
The first term is exactly the SM expression, and the second term is clearly positive provided $M_i>M_W$.

\newsection{New heavy Quarks $\bm{T_i}$}\label{sec:Ti}

We will now assume the existence of an arbitrary number of heavy
singlet quarks $T_i$ with electric charge $+2/3$ that can mix with the
top quark. For simplicity we will suppress the mixing between the
$T_i$'s, but from the arguments given below it will be evident that our
proof is valid also for this more complicated case.

It is well known \cite{Branco} that the mixing of new singlet heavy quarks with the standard quarks implies the presence of FCNC transitions at tree level. Consequently this case, strictly speaking, does not belong to the CMFV framework. However, as far as $K$ and $B$ physics is concerned, the mixing of singlet $T_i$ quarks with the top quark does not imply FCNC contributions at tree level, and thus models of this type belong to the CMFV class if only processes with external down-type quarks are considered.

The gauge-fermion vertices take now the general form $(q=d,s,b)$
\bea
\bar t q W^\mu\quad &:&\quad i\frac{g}{2\sqrt{2}}\gamma^\mu(1-\gamma_5)V_{tq}c_t\\
\bar T_i q W^\mu \quad&:&\quad
i\frac{g}{2\sqrt{2}}\gamma^\mu(1-\gamma_5)V_{tq}s_i
\eea
with $0<c_t<1$ signalling the violation of the $3\times3$ CKM
unitarity. For instance in the LH model we have at $\ord(v^2/f^2)$
\cite{BPU}
\be
c_t=1-\frac{x_L^2}{2}\frac{v^2}{f^2}\,,\qquad s_T=x_L\frac{v}{f}\,,
\ee
where $v=246\gev$ is the SM Higgs VEV, $f\sim 1\tev$ the new physics scale and $0<x_L<1$ a parameter of the LH model.
Generalizing the discussion in \cite{BPU} to an arbitrary number of
$T_i$ fields, the unitarity relation involving also $T_i$ is given by 
\be
\lambda_u^{(q)}+\lambda_c^{(q)}+c_t^2\lambda_t^{(q)}+\sum_i
s_i^2\lambda_t^{(q)}=0
\ee
with the ``bare'' $\lambda_i^{(q)}$ still satisfying
\eqref{eq2.10}. Therefore we also have
\be\label{eq3.5}
c_t^2+\sum_i s_i^2=1\,.
\ee
We will now generalize \eqref{eq2.1} to include the mixing with the
heavy quarks $T_i$.

Including $u$, $c$, $t$, $T_i$ quark contributions to the box diagrams we
find the following linear combination (we suppress the index $q$ of $\lambda_i^{(q)}$) for
each diagram in Fig.~1 that now includes also $T_i$ exchanges:
\bea
\lambda_t^2S(y)&\propto&c_t^4\lambda_t^2 F(t,t)+\lambda_c^2 F(c,c)+\lambda_u^2
F(u,u)+\sum_i s_i^4 \lambda_t^2 F(T_i,T_i)\nn\\
&&+\, 
2c_t^2\lambda_c\lambda_t F(c,t)+ 2c_t^2\lambda_u\lambda_t F(u,t)+ 2\lambda_u\lambda_c F(u,c)+ \sum_{i\ne j} s_i^2s_j^2 \lambda_t^2 F({T_i,T_j})\qquad\nn\\
&&+\,2\sum_i s_i^2
c_t^2 \lambda_t^2 F(T_i,t)+2 \sum_i s_i^2\lambda_u\lambda_t F(T_i,u)+2 \sum_i s_i^2\lambda_c\lambda_t F(T_i,c)\,.
\eea
Using $m_u, m_c\ll m_t$ and $\lambda_u+\lambda_c=-\lambda_t$
gives then
\bea
S(y)&\propto& c_t^4 F(t,t)+F(u,u)-2c_t^2
  F(u,t)+\sum_i s_i^4 F(T_i,T_i)\nn\\
&&+\,\sum_{i\ne j}s_i^2s_j^2 F(T_i,T_j)+2\sum_is_i^2c_t^2
  F(T_i,t)-2\sum_i s_i^2F(T_i,u)\,.\label{eq3.7}
\eea
This linear combination is the generalization of
the GIM structure \eqref{eq2.1} to include the mixing of $t$ with an
arbitrary number of heavy quarks $T_i$. In the case of a single $T$
quark, \eqref{eq3.7} reduces to the linear combination (4.14) in
\cite{BPU}.

Clearly the basic structure of diagrams contributing here is the same
as in Section \ref{sec:bosons}. Consequently it is sufficient to look only at fermion
propagators as in \eqref{eq2.11}. With \eqref{eq3.7} we find now
instead of \eqref{eq2.11}
\bea
\frac{c_t^4}{(r+m_t^2)^2}&+&\frac{1}{r^2}-\frac{2c_t^2}{r(r+m_t^2)}+\sum_i\frac{s_i^4}{(r+m_{T_i}^2)^2}+\sum_{i\ne
  j}\frac{s_i^2 s_j^2}{(r+m_{T_i}^2)(r+m_{T_j}^2)}\nn\\
&&+\,\sum_i \frac{2s_i^2c_t^2}{(r+m_{T_i}^2)(r+m_t^2)}-\sum_i
\frac{2s_i^2}{r(r+m_{T_i}^2)}\nn\\
&=& \left( \frac{1}{r} - \frac{c_t^2}{r+m_t^2} -
  \sum_i\frac{s_i^2}{r+m_{T_i}^2}\right)^2\nn\\
&=& \left( \frac{1}{r} - \frac{1}{r+m_t^2} +
  \sum_i s_i^2 \left[\frac{1}{r+m_t^2}-\frac{1}{r+m_{T_i}^2}\right]\right)^2\,,\label{eq3.8}
\eea
where in the last step we have used the relation \eqref{eq3.5}. The
first two terms represent the SM contribution. The new physics
contribution represented by the remaining terms in \eqref{eq3.8} is
clearly positive provided $m_{T_i}>m_t$.

In the case of scalar exchanges, represented by $D_2$ and $D_3$, the $u$-quark contributions vanish
due to $m_u\simeq 0$ and we find from \eqref{eq3.7}
\be
\left(\frac{c_t^2 m_t^2}{r+m_t^2}+\sum_i \frac{s_i^2m_{T_i}^2}{r+m_{T_i}^2} \right)^2=\left(\frac{m_t^2}{r+m_t^2}+\sum_i s_i^2\frac{r(m_{T_i}^2-m_t^2)}{(r+m_{T_i}^2)(r+m_t^2)} \right)^2
\ee
with the first term representing the SM contribution. Thus again
provided $m_{T_i}>m_t$ the new physics contribution is positive. {We would like to remark that here it is sufficient to consider the SM case $k_{ia}\propto m_i$, as the presence of additional scalars with more general couplings $\alpha_a+\beta_am_i$ simply results in an additional positive contribution, which can straightforwardly be shown following the lines of Section \ref{sec:bosons}.} 

In summary, we have shown that $\Delta S(y)>0$ also in the presence of
a mixing of $t$ with an arbitrary number of $T_i$ fermions, provided
$m_{T_i}>m_t$, as certainly chosen by nature.

\newsection{Are there any additional Contributions in CMFV?}\label{sec:general}

Although we have so far considered contributions from new charged gauge bosons and charged scalars and from additional singlet $T_i$ quarks separately, it is obvious that our proof also holds if both types of new particles are present simultaneously.

Let us now have a look whether we have missed any type of new physics, involving contributions from charged particles, consistent with the CMFV hypothesis as defined in \cite{CMFV} and satisfying the two properties listed in the introduction. In the next section we will investigate what happens if also neutral particles are considered.
\bi
\item First, one could think of introducing additional sequential heavy quark doublets into the theory. This would, however, induce $N\times N$ flavour mixing with $N>3$ and consequently new sources of flavour violation beyond the CKM matrix.
\item 
Second, let us consider additional heavy singlet quarks $B_i$ with charge $-1/3$, as present e.\,g. in grand unified theories based on the group $E_6$. 
These quarks cannot contribute to the box diagrams in Fig.~1, but their mixing with $d,s,b$  generates a small violation of the CKM unitarity and consequently leads to tree level FCNC contributions to $\Delta B=2$ processes. Clearly such processes are outside the CMFV framework. A detailed discussion of this issue can be found in \cite{Branco}. 
\ei
%Thus we have shown that our proof covers all possible CMFV extensions of the SM.

\newsection{Neutral Particle Contributions}\label{sec:neutral}

Until now we have considered only flavour violating charged particle 
exchanges. As we have shown in this case our bound is clearly satisfied.
In the present section we will have a close look at neutral particle contributions. {Such contributions are present in the MSSM, where gluinos and neutralinos can be present in box diagrams.} Such exchanges are also present in models with new sources of 
flavour violation and it is of interest to investigate whether some 
general statements on the sign of their contributions to $\Delta M_{s,d}$ 
can be made by assuming as in the CMFV models that only the operator $\mathcal{Q}_q$ in (\ref{eq:Qq})
is relevant. 
If the relevant flavour violating couplings were equal to the 
CKM couplings, a contribution like this could mimic contributions
present in CMFV models, even if the flavour violation did not have anything to do with the SM Yukawa couplings. An example is provided by the Littlest Higgs 
model with T-parity (LHT) \cite{BBPTUW}
if we choose the mixing matrix $V_{Hd}$, that parametrizes
the interactions of down-quarks and mirror quarks, to be equal to the 
CKM matrix. In doing this one has to make sure that ${\lambda_c^{(q)}}^2$
contributions to $\Delta M_{s,d}$ are made to be negligible by taking 
the mirror fermions of the first two generations to be quasi-degenerate 
in mass.
Otherwise the general structure in (\ref{eq1.3}) would be modified.
Our discussion below of neutral particle contributions is more general 
than the LHT model.

First, $\Delta B=2$ processes could be mediated directly by tree level FCNCs, as in flavour violating $Z'$ models \cite{Zprime}. This corresponds, however, to a new source of flavour violation that has nothing to to with the CKM matrix and CMFV. Even if accidentally the flavour changing couplings of $Z'$ were equal to $\lambda_t^{(q)}$, a straightforward tree diagram calculation shows that also in this case $\Delta S(y)>0$.

If, on the other hand, the neutral gauge bosons are exchanged in box diagrams, in addition to
the usual diagram on the left in Fig.~2 also ``crossed'' box diagrams 
on the right in Fig.~2 contribute.

\begin{figure}
\center{\epsfig{file=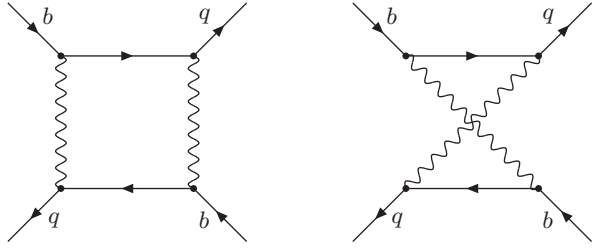}}
\caption{\it Neutral gauge boson contributions to $\Delta B=2$ processes.}
\end{figure}

Let us first consider {the case of} a neutral gauge boson corresponding to a $U(1)$ 
group\footnote{We assume that the possible anomalies related to this 
gauge group in the effective theory are cancelled by new fields in the 
full theory.}. First, as already demonstrated in \cite{BBPTUW,Hubisz} in the 
context of the LHT model, the Goldstone boson contributions cancel 
between the two diagrams in Fig.~2 so that it suffices to consider 
only gauge boson contributions in the 't Hooft-Feynman gauge. Then the 
Dirac structure of the vertices in the box diagrams assures that the 
mass terms in the numerators of the fermion propagators do not contribute.
The first diagram gives then a positive contribution to the function $S(y)$
as in the case of charged gauge boson contributions, while the second 
one, having a negative sign in the four-momentum in the right fermion 
propagator, gives a {\it negative} contribution to $S(y)$. As in the first
diagram 
\be
\gamma_\mu \gamma_\rho \gamma_\nu \otimes \gamma^\nu \gamma^\rho \gamma^\mu 
= 4 \gamma_\mu\otimes \gamma^\mu\,,
\ee
but 
\be
\gamma_\mu \gamma_\rho \gamma_\nu \otimes \gamma^\mu \gamma^\rho \gamma^\nu 
= 16 \gamma_\mu\otimes \gamma^\mu
\ee
in the second diagram, the latter diagram wins this competition and 
we find a strictly negative contribution to the function $S(y)$ that carries
an overall factor $1-4=-3$, which will play some role in a moment.

Thus we found a simple mechanism for a negative contribution to $S(y)$, 
which would be welcome if one could convincingly demonstrate 
that $\Delta M_s < (\Delta M_s)_{\rm SM}$.

Let us now consider the case of a neutral gauge bosons belonging to an adjoint representation of a simple non-abelian group like $SU(N)$ that does not mix with neutral 
gauge bosons corresponding to additional $U(1)$ gauge group factors. As we consider here gauge bosons which can contribute to $S(y)$ only at loop level, the fermions on the internal lines of the box diagrams have to be different from the SM quarks. Such ``mixed'' gauge couplings can only result from the mixing of a new gauge group with the SM $SU(2)_L$ one. This new gauge group thus has to be another $SU(2)$, or an $SU(2)$ subgroup of a larger group. In any way, the only new heavy gauge group which can appear in diagrams with new heavy fermions being exchanged is $SU(2)$. Consequently, we will restrict our attention to this case in the following. 

Now, also the positive charged gauge 
boson contributions have to be taken into account. Moreover, there is one charged, to be called $W_H^\pm$, and one neutral gauge boson, $Z_H^0$, present in $SU(2)$ (note that $W^\pm_H$ has to be counted only once) and the vertices involving the $Z^0_H$ gauge boson carry an  additional suppression factor $1/\sqrt{2}$  relative to the vertices involving the charged gauge boson $W_H^\pm$. 
Thus, the total neutral
gauge boson contribution from the diagrams in Fig.~2 carries an overall factor 
$-3/4$ relative to $+1$ of the charged gauge boson contribution, when the 't Hooft-Feynman gauge is considered. As the charged Goldstone boson 
contributions have been shown in Section \ref{sec:bosons} to be positive and the neutral
ones do not contribute, the total contribution of gauge bosons of a 
simple gauge group including Goldstone bosons  to $S(y)$ is strictly positive.

If the gauge group involves both $SU(2)$  and $U(1)$ 
factors and there is a mixing between neutral gauge bosons belonging 
to different group factors in the process of spontaneous symmetry breakdown,
the situation is clearly more involved and a total negative contribution
cannot be excluded if the number of $U(1)$ factors is sufficiently high and 
the relevant $U(1)$ charges large.
Yet, the neutral gauge bosons have to compete not only with charged 
gauge boson contributions but also with related charged Goldstone 
boson contributions, that generally are more important than gauge
boson contributions if there are heavy fermions in box diagrams. 
Thus also in this more general case it is very likely that the net
contribution to the function $S(y)$ is positive.

An explicit example of this more complicated situation is provided
by the LHT model. For the choice $V_{Hd}=V_\text{CKM}$ one finds that the
$W_H^\pm$ contributions win over the joint $Z^0_H$ and $A^0_H$ contributions
so that in this case $\Delta S(y) >0$, but we cannot exclude that in the 
presence of several $U(1)$ factors $\Delta S(y) < 0$ could be possible.

\begin{figure}
\center{\epsfig{file=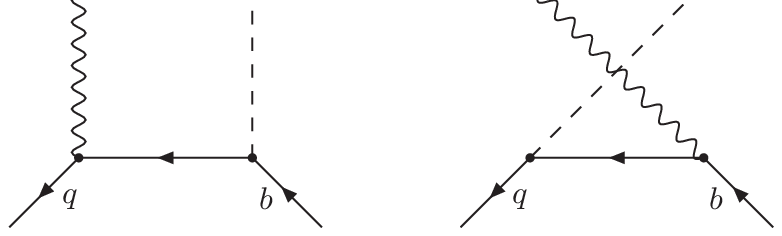}}
\caption{\it Neutral scalar contributions to $\Delta B=2$ processes.}
\end{figure}

 Let us now have a look at physical neutral scalars contributing to $\Delta B=2$ processes. While the ``direct'' and ``crossed'' diagrams involving two scalar exchanges again cancel each other as in the case of neutral Goldstone bosons, this is generally not the case for the mixed gauge boson -- scalar contributions as shown in Fig.~3. The reason is that in contrast to the neutral Goldstone bosons, whose couplings to fermions are always real, the vertices of physical neutral scalars to fermions can in principle be both real and imaginary. In the case of real couplings, the two diagrams in Fig.~3 again cancel each other, because the lower scalar fermion vertices in these two diagrams differ by sign, and we find that such scalars do not contribute to $S(y)$. On the other hand, in the case of imaginary couplings, these vertices are obviously equal, and we thus find that in this case the two diagrams simply add up and bring in a factor of 2. As the first diagram in Fig.~3 has been shown in Section~\ref{sec:bosons} to give a positive contribution to $S(y)$, we finally conclude that physical neutral scalars can only enhance $S(y)$, depending on their couplings  to fermionic fields.

\begin{figure}
\center{\epsfig{file=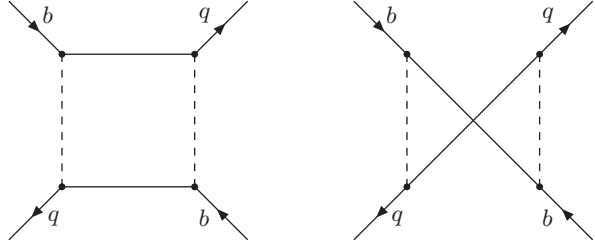}}
\caption{\it Majorana fermion contributions to $\Delta B=2$ processes.}
\end{figure}

Finally, the case of neutral fermion contributions should be considered. In the case of Dirac fermions, which have a specified direction of fermion flow, only the first diagram in Fig.~4 contributes and gives a positive contribution to $S(y)$. In the case of Majorana fermions however,  again direct and crossed diagrams are present, 
as shown in Fig.~4, with positive 
and negative contributions to $S(y)$, respectively. The counting is 
different from the one considered above as now scalars and Majorana fermions instead 
of Dirac fermions and gauge bosons are exchanged. Moreover, the nature 
of Majorana fermions complicates the issue and we were not able to 
reach {a general} clear cut {conclusion} as in the case of gauge bosons and scalars. 
{On the other hand, a recent detailed analysis of the MFV limit of the MSSM at low $\tan\beta$ \cite{ABG} shows that at least in this case {if the higgsino coupling is large compared to the other SUSY scales} the presence of Majorana fermions (gluinos and neutralinos) implies the presence of new operators so that this model, in contrast to the frequently made statements in the literature, does not belong to the CMFV class, and therefore it in principle does not concern our present analysis. However, a remarkable finding of \cite{ABG} is that within the MFV MSSM at low $\tan\beta$ our bound holds even when new operators give non-negligible contributions. This feature is due to an interesting interplay between chargino and gluino box diagrams in the different mass regimes considered in \cite{ABG}. Charginos always give positive contributions, whereas gluino contributions can be negative or positive. However, when the latter are negative, they are always outpaced by large positive chargino contributions, while when the chargino contributions are small, the gluino contributions turn out to be positive.} {Finally, neutralino contributions to $\Delta M_{s,d}$ are always negligible.}

\newsection{Summary}\label{sec:summary}

In this paper we have demonstrated that in any model with CMFV whose low energy limit is given by the SM, the new physics contributions to $\Delta M_{s,d}$ from charged particles are positive, so that $(\Delta M_{s,d})_\text{SM}$ represents the lower bound for $\Delta M_{s,d}$ in this case. 
%Our proof eliminates at the same time the possibility of having in this class of models a fine-tuned scenario where the new physics contributions reverse the sign of $S_0(x_t)$ \cite{BF,mfv2}.

The validity of the bound in question is independent of the scalar masses. As far as the gauge sector is concerned, a sufficient condition for the validity of the bound is $M_i>M_W$, which is satisfied in nature. The same applies to the quark sector, where $m_{T_i}>m_t$ is sufficient to guarantee $\Delta M_{s,d}>(\Delta M_{s,d})_\text{SM}$.

Our discussion of Section 6 shows that in the presence of neutral particle
 exchanges in box diagrams with flavour violating couplings  
equal to the elements of the CKM matrix, our lower bound could in 
principle be violated. Still we have demonstrated that our bound is satisfied for
\bi
\item
neutral scalar particles being exchanged in box diagrams together with neutral gauge bosons,
\item
neutral gauge bosons belonging to $SU(2)$ gauge groups being exchanged in box diagrams with new charged fermions. Their negative contributions to the function $S(y)$ are always overcompensated by the corresponding positive charged gauge boson contributions,
\item
tree level neutral gauge boson contributions like $Z'$.
\ei
On the other hand,
\bi
\item
the contribution of a $U(1)$ gauge boson exchanged in box diagrams is strictly negative. However, if this gauge boson can also contribute at tree level, the total contribution remains positive, 
\item
the contributions of Majorana fermions can have both signs depending on their masses relative to the masses of the scalars exchanged in the box diagrams. {However a detailed analysis \cite{ABG} shows that in the MFV MSSM at low $\tan\beta$ our bound is still satisfied despite the presence of gluino contributions.}
\ei

It should also be emphasized that in a more general formulation of MFV in which new operators in addition to $\mathcal{Q}_q$ are admitted to contribute to $\Heff(\Delta B=2)$ \cite{mfv2}, our bound can be violated, as best illustrated by the MSSM with MFV and large $\tan\beta$ \cite{BCRS}. 

Similarly, in the case of $\Delta F=1$ processes no analogous lower bounds on the short distance functions $X,\,Y,\,Z$ can be derived, so that for instance the sign of $X$ can in principle be reversed within a CMFV scenario, in agreement with the analysis of \cite{BF}.  A well-known example is given by the MSSM with MFV and low $\tan\beta$, where $X$ can be both smaller and larger than $X_\text{SM}$ \cite{MSSMlow,MSSMlow2}. 

It will be of interest to watch the future improved calculations of $F_{B_q} \sqrt{\hat B_{B_q}}$ with the hope that one day we will know whether the lower bound derived here is satisfied in nature. Its violation would imply the relevance of new operators in $\Delta B=2$ transitions and/or the presence of new complex phases beyond the CKM one. A simple example for the latter case is offered by the Littlest Higgs model with T-parity where indeed $\Delta M_{s,d}$ can be smaller than  $(\Delta M_{s,d})_\text{SM}$ \cite{BBPTUW}.
Our discussion of Section \ref{sec:neutral} shows that neutral particle exchanges in box diagrams could also be responsible for the violation of our bound, provided they are not overcompensated by the charged particle contributions in a given model.

\subsection*{Acknowledgements}

We would like to thank Sebastian J{\"a}ger for sharing a related unpublished work with us. 
We also acknowledge useful discussions with Emidio Gabrielli, Gino Isidori
 and 
Andreas Weiler. 
Special thanks are given to Cecilia Tarantino, {Wolfgang Altmannshofer and Diego Guadagnoli} for a careful reading of the manuscript and for  critical comments. {Finally we thank the unknown referee for her/his critical remarks that allowed us to improve our paper.}
This research was partially supported by the German `Bundesministerium f{\"u}r Bildung
und Forschung' under contract 05HT6WOA.

\end{document}